# Polyp detection inside the capsule endoscopy: an approach for power consumption reduction

M. A. Khorsandi[1], N. Karimi[1], S. Samavi[1,2]

[1]Department of Electrical and Computer Engineering, Isfahan University of Technology, Isfahan, Iran
[2]Department of Computational Medicine and Bioinformatics, University of Michigan, Ann Arbor, U.S.A.

**Abstract.** Capsule endoscopy is a novel and non-invasive method for diagnosis, which assists gastroenterologists to monitor the digestive track. Although this new technology has many advantages over the conventional endoscopy, there are weaknesses that limits the usage of this technology. Some weaknesses are due to using small-size batteries. Radio transmitter consumes the largest portion of energy; consequently, a simple way to reduce the power consumption is to reduce the data to be transmitted. Many works are proposed to reduce the amount of data to be transmitted consist of specific compression methods and reduction in video resolution and frame rate. We proposed a system inside the capsule for detecting informative frames and sending these frames instead of several non-informative frames. In this work, we specifically focused on hardware friendly algorithm (with capability of parallelism and pipeline) for implementation of polyp detection. Two features of positive contrast and customized edges of polyps are exploited to define whether the frame consists of polyp or not. The proposed method is devoid of complex and iterative structure to save power and reduce the response time. Experimental results indicate acceptable rate of detection of our work.

## 1      Introduction

Wireless capsule endoscopy (WCE) is a state of the art invention in the field of medical imaging which is introduced in [1]. Painless diagnosis and the ability of capturing frames in inaccessible spots in the digestive tract are most important advantages of this technology over the traditional endoscopy. WCE has a compact structure consist of a camera, an image-processing chip, a battery and a radio transmitter. After activation, patient swallows the WCE and it begins capturing videos from the digestive tract and sending frames by radio transmitter simultaneously.  During the procedure of capturing, the patients are able to continue their normal life. The process of capturing the frames lasts about 8 hours and finishes when the battery runs out. If acceptable frames were captured, they would result in better diagnosis and plays an important role for physicians to choose the appropriate therapy such as surgery, chemotherapy or polypectomy (in case of polyp detection).
Although WCE has many benefits, this new invention has some limitations that its problems have not been solved yet. MicroCam is a commercial type of WCE which its recent version captures just three frames per second with the resolution of 320×320 pixels [2]. This product has a passive movement without human manipulation and it has only the ability of diagnosis. Some medical researches concentrated on weaknesses of WCEs. The work in [3] evaluated the WCE's capturing duration and its effect on diagnosis process. Authors in [3] observed that increasing the capturing time of WCE results in better diagnosis in some cases. In [4], the authors discussed that the quality and number of images per second are important factors for detection of abnormalities. They also mentioned that improvement of imaging and resolution must be considered for the future development of WCEs. Apart from weak imaging, lack of mechanical actuation is considered as another type of WCE's limitations that are discussed in [5]. The work in [5] reviewed some researches for exploiting mechanical actuation for drug delivery and endoscopy surgery, but none of those works has been commercialized yet.

The most important reason of these weaknesses of WCE is its limited power source [3], [4]. There is a limitation in capsule size and consequently a limitation in its battery size that results in less energy storage space. The camera captures large amount of data and transmits them; the process of data transmission consumes large portion of energy. In case of mechanical actuation, the battery must supply the mechanical parts as well, whereas mechanical parts occupies lots of capsule's space [5]. A battery with large capability of energy storage is required to overcome all of problems arose by power source. Nevertheless, technology has not reached a level in which a battery with these capabilities can be manufactured.

Many works are proposed to cope with the problem of power by reducing the data value to be transmitted by performing compression. The work in [6] proposed an improved JPEG-LS encoder to implement a near-lossless compression. The key structure of this design to perform better compression is to use gradient predictor rather than conventional median predictor in JPEG-LS. The work in [7] introduces an architecture for low complex 1D discrete cosine transform (DCT) which is based on Cordic-Loeffler technique. In this architecture a new type of carry-look-ahead and carry-save adder is exploited to reduce the number of shift and add operations, which results in power consumption reduction in compressor chip. Another approach for image compression considers characteristics of WCE images and converting images to a color space apart from RGB. The work in [8] converts image into YUV space, which this conversion results in narrower histogram for WCE images. Then A Golomb-Rice encoder is exploited to perform an excellent compression for this narrow histogram.

Another problem that physicians face while using WCE is its time-consuming review time. Considering eight hours of capturing with three frames per second, about 80,000 frames must be reviewed. Unfortunately, most of these frames are consist of non-informative or redundant data. Many researches are done to cope with this problem outside the capsule and when its frames are sent. They perform their algorithms on received data, detect non-redundant or informative frames, and drop the other frames. The work in [9] performed a method to detect similar frames with respect to small movement of capsule for consecutive frames. For this purpose, three approaches based on intensity, features and motion from optical flow is exploited to reach the frame similarity. Another work in [10] proposed a method to detect specific abnormalities consist of ulcers and polyps. The intensity and Hue of input image is considered to detect polyps and ulcers respectively, then a log Gabor filter is applied for segmenting and detecting respective abnormality. The work in [11] performed saliency model for video summarization. The gradient, contrast feature and frame similarity are considered as features to detect salient frames. However, all of these mentioned works perform their algorithms outside the capsule. In our previous work in [12] we performed a simple and fast frame similarity detection method based on moments of image to prevent similar frames to be sent. The proposed system was implemented as an architecture on an FPGA to prove its capability of implementation inside the WCE. Hence, a solution for both power source and review time problems of WCE is suggested.

In this paper we proposed a real time structure to detect abnormality inside the WCE. Detection and transmission of abnormal frames will result in power consumption reduction. Polyps are known as important abnormality in small bowl and our work is concentrated on polyp detection. We considered two features of positive contrast and customized edges to detect frames that they consist of polyp. To perform faster detection, we take advantages of integral image for filtering and applying positive contrast feature. The proposed structure is simple and the time between image entrance and processing data is short. The experimental results show the acceptable detection ratio of our method.

The rest of this paper is organized as follow: In section 2, we discussed specifications of polyps which are exploited in our work. In section 3, the proposed method consists of preprocessing, positive contrast feature, customized edge and fusion phase is presented. Section 4 consists of experimental results to verify our work. The final section is dedicated to conclusion.

## 2 Polyps' Specifications

In this work, we focused on polyp detection as one of the common abnormalities in small bowl. The polyps are not considered as a dangerous disorder in body unless it is treated on time. There is a possibility of converting polyps to tumors, so WCE as an important diagnosis tool is used to detect the polyps in human body. After verification of polyp existence, the appropriate treatment such as surgery or polypectomy must be done.

To perform a structure of polyp detection, we considered some physical features of polyps to assist us in our detection algorithm. Polyps are growing tissues that have an inflamed shape. In general, polyps have no specific color specification and their color is the same as neighboring area. The convex shape of polyps is an important feature in our work. In case of uniform light radiation, detection of a convex shape is a challenging issue. Fortunately, due to centralized radiation of light from WCE, detection of polyps is feasible. When centralized light radiates to a polyp, the beams are reflected due to the angle of surface. This concept is shown in Fig. 1a.

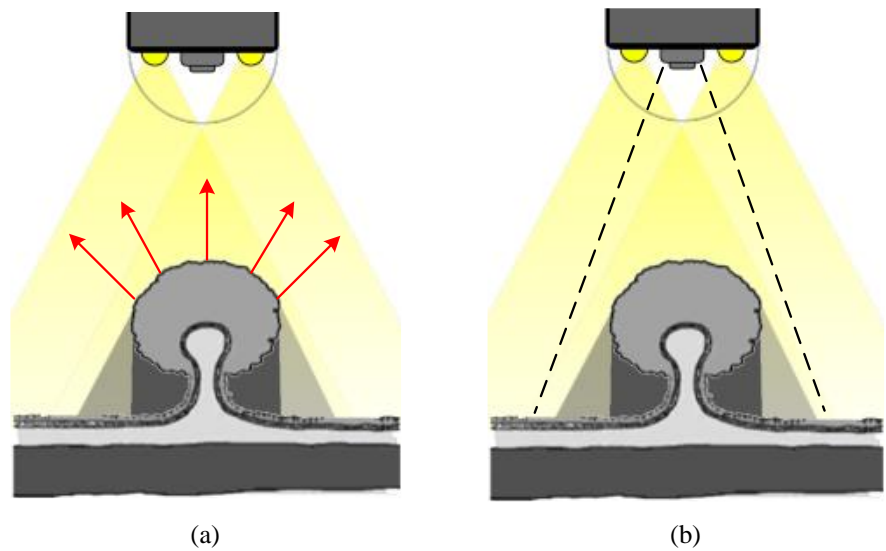

**Fig. 1.** The effect of WCE's light sources to polyps. (a) The radiated light from WCE and reflected light direction. (b) The field of view of camera that includes spots in which the polyp casts shade on it.

The camera receives the reflected light but the reflection from the center of polyp is received considerably rather than sides due to the reflection angle. Thus, the center of polyp seems brighter than sides. On the other hand, the polyps are closer than their adjacent small bowl wall to the WCE. The inflamed shape of polyp leads to creation of this space which results in brighter polyp and darker bottom wall.

Another feature is obtained by the placement of LEDs; The LEDs are placed on the 4 sides of camera. Due to this placement, polyp prevents light to radiate some spots, but these spots are in the field of view of camera. The multisource light from LEDs causes the polyp to cast a shade on the wall of small bowl. The LED on the left of camera causes a shade on the right side of polyp and the shade on the left side of polyp is resulted from the right LED's radiation (usually WCEs have 4 LEDs, but we simplified this concept by considering 2 LEDs for better understanding and drawing). This concept is shown in Fig.1 b. Although one LED lights each shade, the further distance of small bowl walls and lack of the other LED's light result in a narrow dark area between polyp and its bottom. Considering these points in WCE's images, polyps' center is brighter than their sides; Polyps have sharp edges which these edges have a small value of intensity (dark). Two examples of polyps and their related intensity color map is depicted in Fig. 2.

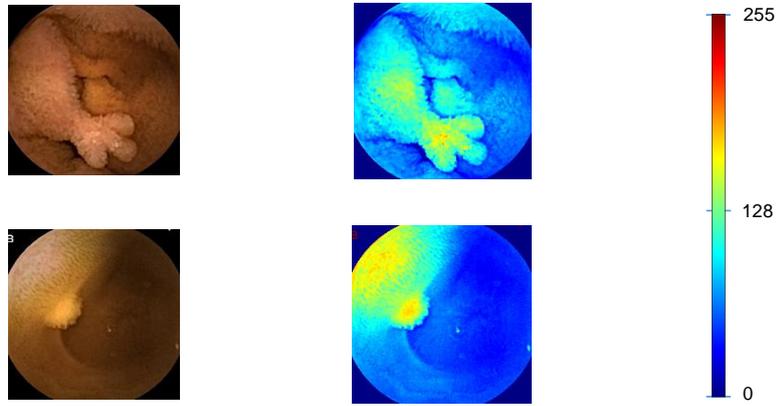

**Fig. 2.** Two sample images of polyps captured by WCE and their related intensity color map. Polyps are brighter in center and darker at sides; also, they have sharp edges with a small value of intensity.

## 3  Proposed method

### 3.1  System Overview

In the first phase of the algorithm, the intensity value of input image is obtained and the integral image (a structure to speed up subsequent phases) is generated from the intensity. These two steps are preprocessing phases for extracting desirable features. Positive contrast feature and customized edges are extracted from integral image. These two mentioned features are not able to represent a polyp completely. For this purpose, we have a fusion phase that fused these features and extracts approximate region of polyp. The system overview of our work is shown in Fig. 3.

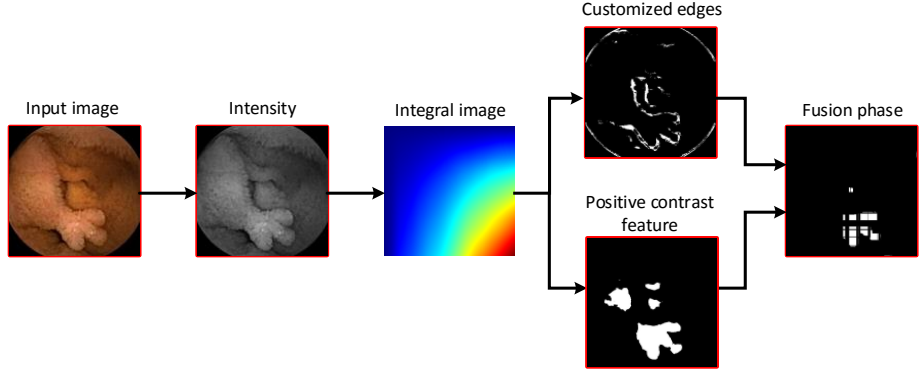

**Fig. 3.** The system overview of proposed method

### 3.2 Preprocessing Phase

Preprocessing phase is responsible to prepare data for subsequent phases. First, the intensity value of input RGB image is calculated. The reason of this conversion is due to lack of color feature in polyps. Therefore, the RGB data or any other color space is not useful but intensity. The next step in the preprocessing phase is generation of integral image. Integral image is introduced in [13] as a structure which is mainly used to speed up calculation of sum of any rectangular area in the image. By taking advantages of this structure, instead of accessing all pixels in an area, just four corner pixels in integral image is read to calculate the sum of rectangular area. Considering $Int\_Im$ as integral image and $Im$ as the image pixel, the integral image is calculated by (1) and by considering a rectangular area between (X0,Y0) and (X1,Y1) the sum of this area is calculated by (2).

$$Int\_Im(Xm, Ym) = \sum_{i=1}^{X_m} \sum_{j=1}^{Y_m} Im(i,j) \quad (1)$$

$$\sum_{i=X_0}^{X_1} \sum_{j=Y_1}^{Y_1} Im(i,j) = Int\_Im(X0, Y0) - Int\_Im(X1, Y0) - Int\_Im(X0, Y1) + Int\_Im(X1, Y1) \quad (2)$$

Although integral image increases the speed of subsequent processes, the process of integral image generation is complex and time consuming. To cope with this problem, some works are proposed to speed up and reduce the power consumption of this process [14], [15]. These works propose architecture for implementation of integral image generation using pipeline and parallel structures to achieve faster performance.

### 3.3 Positive contrast feature

As we discussed before, the intensity value of polyps is larger than its surrounding area. To exploit this characteristic, we introduced positive contrast feature and the goal is to generate a positive contrast mask (PCM). The related equation of PCM is indicated in (3).

$$PCM(Xm, Ym) = \begin{cases} 1 & if\ (\mu_a > \mu_b \times 1.25) \\ 0 & else \end{cases}$$

Where PCM is a binary mask with a size the same as input image and $\mu_a$ and $\mu_b$ are the mean value of pixel in a square size of a and b respectively. The shape of used structuring element to obtain PCM is shown in Fig. 4.

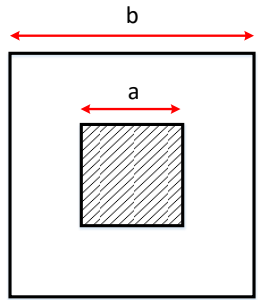

**Fig. 4.** The used structuring element for producing PCM

Computation of $\mu_a$ (and similarly $\mu_b$) is simplified by using integral image as indicated in (4).

$$\mu_a = \frac{\sum_{i=X_m-\frac{a}{2}}^{X_m+\frac{a}{2}} \sum_{j=Y_m-\frac{a}{2}}^{Y_m+\frac{a}{2}} Im(i,j)}{a \times a} \quad (4)$$

This structuring element slides on the image and the condition of positive contrast feature is checked. When this condition is met, the location of the pixel at the center of structuring element is considered and a bit in the PCM at the same address is set to one.

Each non-zero bit in PCM indicates a bright area surrounded by a dark area. It seems that instead of small square (a), we could have used only the intensity of central pixel. But the role of this square to perform a smoothing filter to remove sporadic changes of intensity.

### 3.4 Customize Edges

In section 3, we discussed specifications of polyps' edges. In our work, we exploited gradient as a feature for detection of polyps' edges. Considering polyps' specifications and human digestive track environment, the gradient must be supplemented by other inputs to detect polyps' edges. These edges are sharp and dark, thus only gradients with high differential value and low level of intensity must be considered.
By considering these specifications, final edges are formulated in (4).

$$E(i) = \begin{cases} 1 & (im(i) > im(j) + \tau_1) \text{ and } (im(j) < \tau_2)) \\ 0 & else \end{cases}$$

$im(i)$ indicates the image pixel intensity value and $im(j)$ indicates the adjacent pixels in 4-connected neighbors (the pixels on up, down, left or right side). There are 2 conditions with 2 threshold value for this mapping which they are in accordance with the specifications of polyps. $\tau_1$ indicates the distance between intensity of the pixel and its adjacent pixel. Also, the adjacent pixel must have an intensity value less than $\tau_2$.

The horizontal and vertical edges are extracted and the direction of edges are obtained as well. The direction of edge indicates the darker and brighter sides of edge. It is clear that a polyp must have an edge in which the polyp is located in the brighter side of edge.

### 3.5 Fusion Phase

The final phase of polyp detection is fusion phase, which is responsible to extract polyp's approximate region. The positive contrast feature and customized edges extract some features that are related to polyps; unfortunately, there is a possibility that some other areas are extracted as well. Hence, there must be a method to fuse these two features in a way that only polyps are extracted.

We know that polyps have a round and closed shape. At first glance, it seems that polyps are detected if they have a positive contrast feature and the edges must surround them. However, due to position of WCE during capturing and the direction of polyp, the edges might not surround the polyps completely. Therefore, we consider only two perpendiculars edges as sufficient inputs for our problem. So a polyp is detected if it has a positive contrast feature and there are at least two perpendiculars edges, which their brighter side are on the polyp.

The details of this algorithm is as follow: There are two horizontal and vertical masks that they are used to indicate acceptable fusion of edges and PCM in horizontal and vertical directions. The PCM is scanned from four different directions separately (left to right, right to left, up to down and down to up). When a non-zero value is detected, it is checked if there is an edge in its direction or not. For example, if the PCM is scanned from left to right, an edge that its dark side is on the left and the bright side is on the right is considered. When these conditions are met, the pixels on the related mask (in this example horizontal mask due to left to right scan) are set to one and the scan is continued until a zero PCM pixel is found. After applying this algorithm for four directions, a logical and operator is applied to horizontal and vertical masks to obtain the final result. These concepts are shown in Fig. 5.

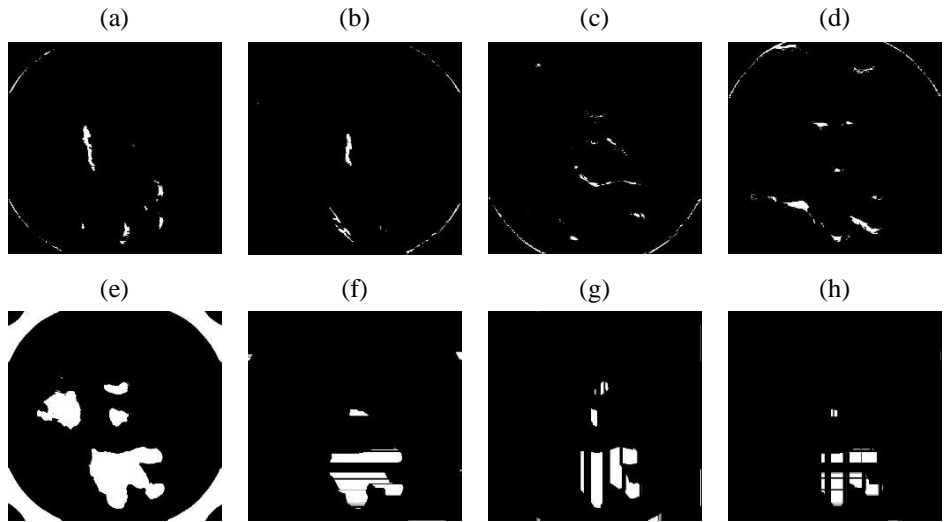

**Fig. 5.** An example of final fusion phase. (a) Right to left edge. (b) Left to right edge. (c) Up to down edge. (d) Down to up edge. (e) Positive contrast mask. (f) Horizontal mask. (g) Vertical mask. (h) Final mask.

## 4   Experimental results

Unfortunately, there is not any published WCE dataset for researchers and almost all of works in this field (consist of [9] and [10]) have not used a common dataset and consequently no direct comparison with other works is available. To verify our work, we tested our algorithm with several videos downloaded at WCE official website [16]. These videos consist of both informative and non-informative frames. As we discussed before, the proposed method detects informative frames specifically frames which consist of polyp inside the WCE. Although the proposed method detects approximate region of polyps, our goal is not to segment polyp's region. The suspicious frames are sent and then physicians or a sophisticated computer program can review and evaluate the received frames.

We have to define a parameter to determine whether the frame consist of polyp or not. For this purpose, we consider a threshold value on the number of pixels in the final mask which is selected 500. Other threshold values in our work is as follow:
- Size of filter for positive contrast feature: a=16, b=64.
- Thresholds for edge detector: $\tau_1 = 2$, $\tau_2 = 100$.

In Table 1, the accuracy of our work is shown by considering these threshold values.

**Table 1.** Accuracy of our work

| Real<br>Our result | Informative (true) | None-informative (false) |
|---|---|---|
| İnformative (positive) | 0.81 | 0.12 |
| None-informative (negative) | 0.19 | 0.88 |

The results of three frames that are correctly labeled as informative and three frames that are correctly labeled as non-informative are shown in Fig. 6 and Fig. 7 respectively.

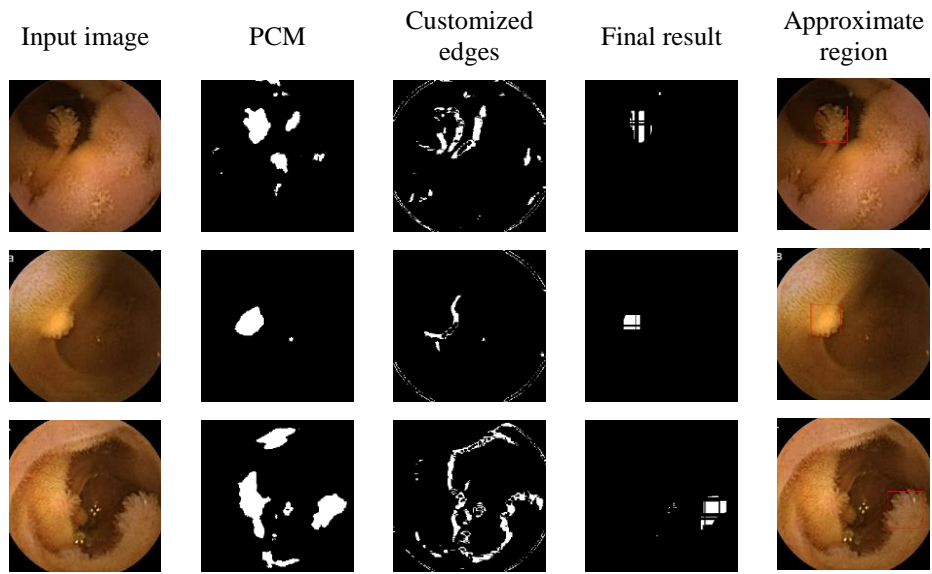

Fig. 6. Three frames consist of polyp and related PCM, edge and final fusion.

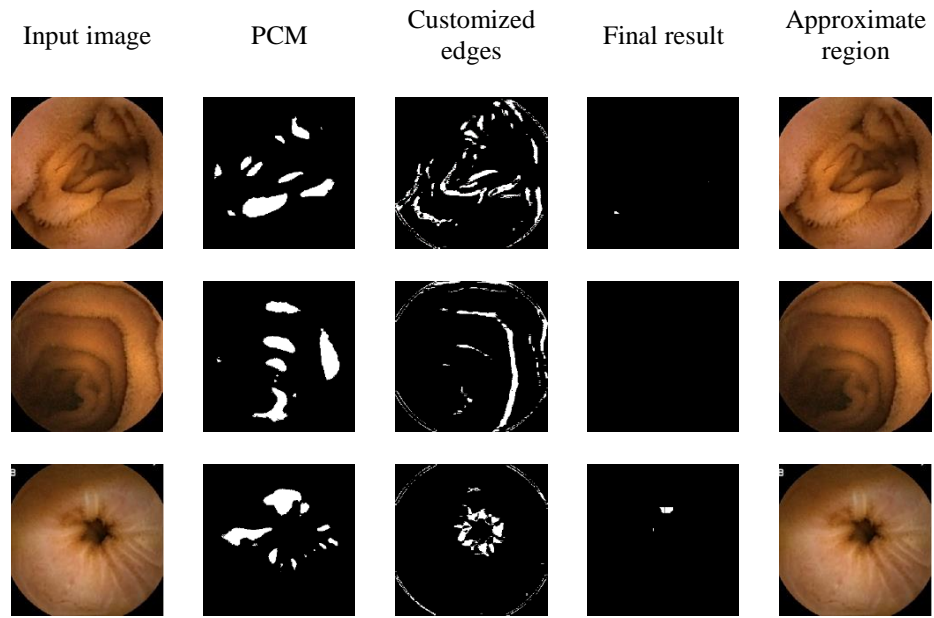

Fig. 7.   Three normal frames and related PCM, edge and final fusion.

## Conclusion

In this paper, we proposed an algorithm for performing polyp detection. We considered polyps' geometry as convex and inflamed structures in human digestive tract. Then two features of positive contrast feature and customized edges are extracted and then fused to obtain results. The key elements of our algorithm is based on directional gradient and integral image which are both able to be implemented in parallel and using pipeline in their structure. On the other hand, this algorithm is not complex and devoid of any floating point calculations. These specifications result in possibility of implementation of our algorithm on hardware and inside the capsule rather than other methods, which are implemented outside the capsule.